\documentclass{article}
\usepackage{ijcai21}
\usepackage{xcolor}

\usepackage{times}
\usepackage{soul}
\usepackage{url}
\usepackage[hidelinks]{hyperref}
\usepackage[utf8]{inputenc}
\usepackage[small]{caption}
\usepackage{graphicx}
\usepackage{amsmath}
\usepackage{amsthm}
\usepackage{booktabs}
\usepackage{algorithm}
\usepackage{algorithmic}
\title{Active Learning Meets Optimized Item Selection
}

\author{
Bernard Kleynhans$^1$\footnote{Contact Author}\and
Xin Wang$^1$\and
Serdar Kad{\i}o\u{g}lu$^1$\\
\affiliations
$^1$AI Center of Excellence \\ Fidelity Investments, Boston, USA\\
\emails{firstname.lastname@fmr.com}
}

\begin{document}

\maketitle

\begin{abstract}
 Designing recommendation systems with limited or no available training data remains a challenge. To that end, a new combinatorial optimization problem is formulated to generate optimized item selection for experimentation with the goal to shorten the time for collecting randomized training data. We first present an overview of the optimized item selection problem and a multi-level optimization framework to solve it. The approach integrates techniques from discrete optimization, unsupervised clustering, and latent text embeddings. We then discuss how to incorporate optimized item selection with active learning as part of randomized exploration in an ongoing fashion.
    
    
    
\end{abstract}

\vspace{-0.25cm}
\section{Introduction}

Recommender systems have become central in our daily lives and are widely employed to help users discover relevant content. The classical setting is composed of a set of users, $U$, and a set of items, $I$, from which top-$k$ items are chosen and shown to the user. 

In this setting, notice there is an apriori decision to determine the \textit{universe of items} $I$ that can be recommended. We refer to this combinatorial problem as the Item Selection Problem (ISP). In our recent work~\cite{kadioglu2021ISP}, we presented a multi-level optimization approach for selecting items to be included in experimentation. The main goal of this approach is to minimize the cardinality of the item universe while maximizing item diversity. By minimizing the cardinality, we reduce the experimentation time window and mitigate undesired user experience and business impacts while the recommender system is collecting the necessary training data to build personalization models. To that end, we show how to use a latent embedding space to calculate diversity measures between items and maximize the diversity of the selected items. Using the item embedding we propose a simple warm-start procedure to enable transfer learning from the \textit{randomized exploration} phase to the \textit{personalized exploitation} phase. More broadly, our hybrid approach stands out as an integration block between modern recommender systems and classical discrete optimization techniques.

\smallskip

In this paper, we start with an overview of the problem setting (Section 2) and then present our multi-level optimization framework (Section 3) based on~\cite{kadioglu2021ISP}. We then show how to apply transfer learning via warm-start (Section 4) to increase personalization capacity. The work outlined so far considers solving ISP once at the inception of the system. Our main contribution is a proposal to take this approach a step further by incorporating active learning (Section 5) for effective experimentation and exploration in a continuous fashion.

\begin{figure}[t]
\begin{center}
    \includegraphics[height=1.5in, width=3in]{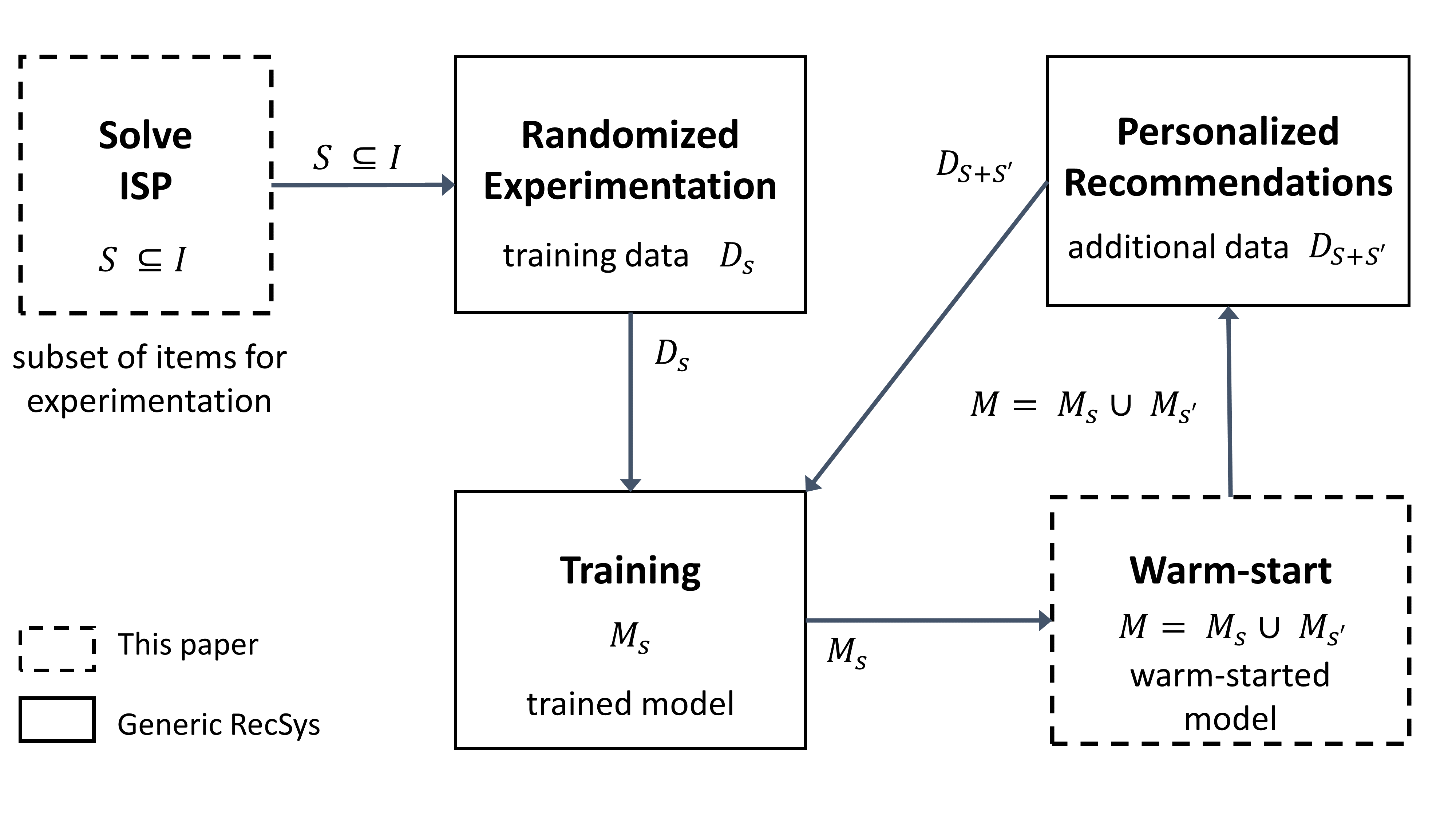}
\end{center}
\vspace{-0.2cm}
\caption{Item selection to speed-up experimentation for personalized recommendations.}
\label{fig:ISP}
\vspace{-0.3cm}
\end{figure}

\section{Problem Definition}

\noindent\textbf{Item Selection Problem (ISP):} Given a set of items $I$, the goal of the ISP is to find the minimum subset $S \subseteq I$ that covers a set of labels $L_c$ within each category $c \in C$ while maximizing the diversity of the selection $S$ in the latent embedding space of items $E(I)$.

\smallskip

\noindent{\textbf{Illustrative Example:}} Consider for example a movie recommendation system. In a movie recommender system, the \textit{items} $I$ correspond to all available \textit{movie titles} that could be recommended. The \textit{categories} of interest, $C$ can include the genre and language for example. Within each category $c \in C$, we can have a set of \textit{labels} for genre (e.g., action, comedy, thriller) and  language (e.g., English and French). The ISP seeks to include at least one movie from each label $L_c$ for different categories $c \in C$, while maximizing the diversity of selected movies in the \textit{latent embedding space} $E(I)$. The latent representation can be based on textual data (e.g., synopses, movie reviews) or image data (e.g., cover art). 

\vspace{0.5cm}

The ISP is most relevant when there exists limited or no historical data. As illustrated in Figure~\ref{fig:ISP}, randomized experimentation is employed to collect training
data $D_S$ that is used to build personalization models $M_S$. The longer the exploration phase takes, the worse user experience and business outcomes are.
To mitigate this, our strategy focuses on solving the ISP to guide randomized exploration which is later augmented with warm-started models $M_{S^{'}}$.  

\section{Discrete Optimization to Solve the ISP}

Our approach to solving the ISP is closely related to the classical Set Covering Problem (SCP)~\cite{beasley1987algorithm} which we embed in a multi-level optimization framework. It consists of three levels; finding the minimum subset size, maximizing diversity and maximizing coverage within a fixed bound. 

\smallskip

\noindent\textbf{Minimum subset size:} Let $P_{unicost}$ be a standard covering formulation to select a subset of items that cover all predefined labels. Assume $unicost\_selection \subseteq I$ is the solution with $k$ number of selected items.   

\smallskip

\noindent\textbf{Maximizing diversity:} Given the minimum subset size $k$ from the solution of $P_{unicost}$, we cluster the embedding space of items $E(I)$ into $k$ clusters and let $K$ denote the cluster centers. We reformulate $P_{unicost}$ by changing its cost structure such that the inclusion of item $i$ incurs cost, $c_i$, based on the distance to its closest cluster. The solution of $P_{diverse}$, denoted by $diverse\_selection$, is the minimum subset of items that are most spread out from each other in the embedding space $E(I)$ while still covering all predefined labels. 

\smallskip

\noindent\textbf{Bounded subset size:} Given $t \leq |P_{diverse}|$ we select up to $t$ items from $diverse\_selection$ such that coverage is maximized and refer to this formulation as $P_{max\_cover@t}$.

\smallskip

Bringing these components together, Algorithm 1 depicts our multi-level optimization framework that consists of solving $P_{unicost}$, $P_{diversity}$ and $P_{max\_cover@t}$. For more details and the problem formulations, we refer to our recent work~\cite{kadioglu2021ISP}.

\begin{algorithm}[t]
\begin{algorithmic}
\STATE {\bf Multi-Level Optimization for ISP(I, M, E, t)}
\STATE \textbf{In:} Items: $I$  
\STATE \textbf{In:} Incident Matrix: $M[label][item]$  
\STATE \textbf{In:} Embedding Space: $E(I)$
\STATE \textbf{In:} Maximum Subset Size: $t$
\STATE \textbf{Out:} Selected Items: $S \subseteq I$ 
\item[]

\STATE // First Level: Minimize the subset size
\STATE \textbf{Formulate} $P_{unicost}(I, M)$ 
\STATE $unicost\_selection$ $\leftarrow$ \textbf{solve}$(P_{unicost})$

\item[]
\STATE // Second Level: Maximize diversity
\STATE $k$ $\leftarrow$  $ \lvert unicost\_selection \rvert$ 
\STATE $K$ $\leftarrow$ \textbf{cluster}$(E(I), num\_clusters=k)$ 
\STATE \textbf{Initialize} $cost$ $\leftarrow$ zeros$(\lvert I \rvert)$
\FORALL{item $\in$ $I$}
    \STATE $cost_{item}$ $\leftarrow$ \textbf{min}(\textbf{distance}$(item, centroids \in K))$
\ENDFOR
\STATE \textbf{Formulate} $P_{diverse}(I, M, cost, unicost\_selection)$ 
\STATE $diverse\_selection$ $\leftarrow$ \textbf{solve}$(P_{diverse})$

\item[]
\STATE // Third Level: Maximize bounded coverage
\STATE $t$ $\leftarrow$ $ \lvert diversity\_selection \rvert$
\STATE \textbf{Formulate} $P_{max\_cover@t}(diverse\_selection, M, t)$ 
\STATE $S=max\_coverage$ $\leftarrow$ \textbf{solve}$(P_{max\_cover@t})$

\item[]
\STATE \textbf{return} $S$
\end{algorithmic}
\caption{Multi-Level Optimization for ISP}
\label{algo}
\end{algorithm}

\section{Latent Representations for Warm-start}\label{warm-start}

Given the solution from ISP, the experimentation phase can start, which yields training data $D_S$ that is used to build personalization model $M_S$. As shown in Figure~\ref{fig:ISP}, we can use transfer learning~\cite{DBLP:conf/icml/CaruanaNCK04} to expand the capacity of personalization via warm-start. We propose a procedure to warm-start items $s' \in S':I\setminus S$ to build $M_{S'}$ by sharing knowledge from $M_S$. We take advantage of the latent embedding space $E(I)$ to compute pairwise distances between items and find the closest item $s \in S$ for each \textit{untrained} item $s'$. To use $s$ for the warm-start of $s'$, we enforce $distance(s, s') \le w$ for $w > 0$ to ensure that the items are sufficiently similar. We obtain the distance threshold $w$ from the distribution of pairwise distances within a specified quantile $q$. Notice how this allows us to dynamically set the threshold $w$ for the data at hand without requiring a tuning process. For transfer learning between $s$ and $s'$, we leverage the training data $D_s$ or trained parameters of model $M_s$.

\section{ISP + Active learning}





The key idea behind Active Learning (AL)~\cite{settles2009active} is that a machine learning algorithm can achieve greater accuracy with fewer labeled training instances if it is allowed to choose the training data from which it learns. In the context of recommender systems, this is accomplished by letting the system influence which items a user is exposed to in order to learn users' preferences~\cite{rubens2015active} more efficiently. In this section, we outline how the ISP can be incorporated in an active learning framework to continuously inform randomized exploration.

\smallskip

For an operational recommender system to work effectively in a dynamic environment, we need to solve two challenging problems: initially, on Day\textendash0, a large number of items have no or inadequate feedback data. Subsequently, on Day\textendash1+, new items are periodically added to the system with no historical training data to learn from. Solving the ISP, as outlined earlier, addresses the first problem on Day\textendash0. It provides an offline selection that helps minimize the cardinality of the item universe while maximizing the diversity of items. 

\smallskip

The important realization is that there is an ongoing need for experimentation for a recommender system to continue to explore users' behaviors and interests. In fact, the continued exploration is a more challenging task than solving the one-shot ISP. In the beginning, all items are candidates for the initial ISP selection. However, as time progresses, the system needs to distinguish between items that are trained effectively vs. items for which further engagement data is still necessary. 

Recommender systems need to balance exploration and exploitation in some form. Hence, the ISP continues to be relevant beyond Day\textendash0, and value can be gained by solving it in an ongoing fashion and incorporating its solution dynamically as the system proceeds.

\smallskip

As part of our ongoing work, let us share initial directions on how the ISP can be further integrated into the exploration component of a recommender system. The main goal of this line of research is to adopt an exploration strategy that is not purely random but provides a principled approach that can be tailored to specific criteria. For instance, beyond improving the accuracy of recommendations, we can aim to explore items to warm-start the maximum number of cold items. 


 
 \smallskip
 
\begin{figure}[t]
\begin{center}
    \includegraphics[height=1.5in, width=3in]{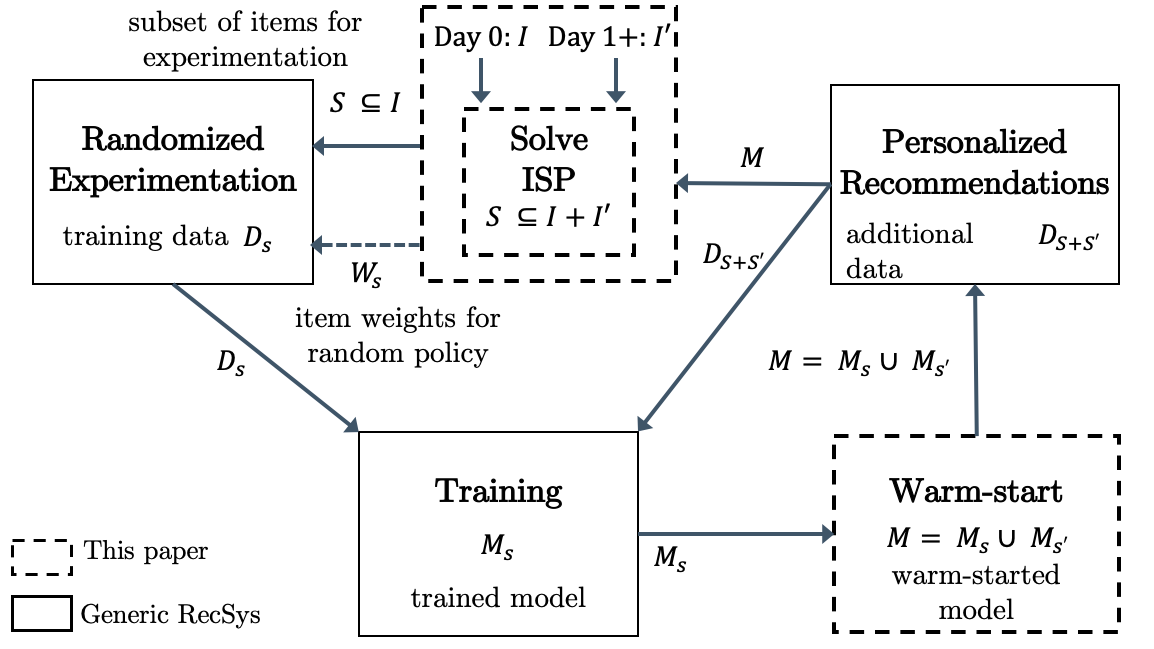}
\end{center}
\caption{ISP with active learning in exploration.}
\label{fig:ISP+AL}
\vspace{-0.5cm}
\end{figure}

Figure~\ref{fig:ISP+AL} illustrates our initial design on how to perform active exploration by integrating ISP within the recommender system. In contrast to Figure~\ref{fig:ISP}, instead of performing a once-off item selection, the idea is to solve for which items to include in the exploration continuously. As shown in Algorithm~\ref{algo}, the multi-level optimization algorithm for solving the ISP can be applied recursively to select a fixed number of items from the set of remaining cold (untrained) items without replacement until all items are sufficiently trained. 

\smallskip

Another exploration policy is to select items from iterations until all are selected and at each iteration with ISP, items are assigned a weight based on the order in which it was selected. The earlier selected items receive a higher weight. Using these weights, we can generate recommendations in the exploration stage that favor items yielding better coverage of item labels and are more diverse as guaranteed by the provably optimal ISP solution.

\smallskip

Alternative exploration policies could also be readily implemented with minor modifications to the ISP. For example, we could assign the highest weight to items that greedily warm-start the most number of cold items.



\smallskip
Moreover, the active exploration component can be made interactive with human inputs, letting the procedure become an instance of human-in-loop optimization. Besides coverage and diversity, the active exploration can be further guided to cover other preferences, such as time-sensitive or seasonal information. For instance, in an e-commerce application, a certain category of seasonal items can be weighted more in the exploration for a fixed period of time. Similarly, newly released products with large potential to promote sales could be of more interest from a business perspective. Such constraints (preferences) can be incorporated into our ISP formulation as side-constraints. 

\vspace{-0.25cm}
\section{Numerical Results for ISP}

Let us first provide a summary of our numerical results that demonstrate the speed-up enabled by solving the ISP in the randomization phase while ensuring diversity and transfer learning capacity. We then conclude with an evaluation protocol that can be used to evaluate the effectiveness of combining ISP in the Active Learning process as suggested in Section 5. 

\smallskip

\noindent\textbf{Datasets:} We use the Goodreads Book Reviews~\cite{DBLP:conf/recsys/WanM18} dataset with 11,123 books and the MovieLens (ml-25m) Movie Recommendations~\cite{movielens2015} dataset with 62,423 movies. We consider two randomly selected subsets, small and large versions with 1,000 and 10,000 items, respectively. We are interested in selecting movies (books) for exploration that cover all (or maximum) genres, producers (publishers), languages, and genre-language (genre-publisher) combinations. Detailed descriptions of the datasets can be found in~\cite{kadioglu2021ISP}.

\smallskip

\noindent\textbf{Analysis of coverage:} To find the minimum set of items covering all labels, we solve $P_{unicost}$ and compare the number of selected items and the resulting label coverage between our approach and several challenger algorithms. We show that all labels can be covered by selecting a fraction of items thereby reducing exploration time. Compared to $Random$ and $KMeans$ based challengers the set covering formulations are shown to provide substantially better coverage. $P_{unicost}$ achieves complete coverage by selecting between 5\% and 37\% of available items. For the same number of items, $Random$ and $KMeans$ can only achieve coverage of between 30\% and 45\%.

\smallskip

\noindent\textbf{Analysis of bounded coverage:} To further demonstrate potential speed-up in random experimentation, we vary the subset bound $t$ and analyze the label coverage before and after warm-starts for $P_{max\_cover@t}$, $KMeans$ and $Random$. Critically we show that for a given coverage level, the required number of items $t$ is always lower for $P_{max\_cover@t}$ and that for the same bounded $t$, it yields coverage 2-4X higher than challenger methods. After the warm-start, coverage increases for each method at each t, and notably, the coverage for $P_{max\_cover@t}$ continues to rank highest.

\smallskip

\noindent\textbf{Analysis of warm-start:} To assess the effectiveness of the warm-start procedure we perform sensitivity analysis on the distance quantile $q$ and evaluate the average number of labels covered per item (unit coverage) after warm-start. We show that $P_{max\_cover@t}$ consistently has the highest unit coverage and performs significantly better than $Random$ and $KMeans$ for top (semi-) deciles, i.e., $q\leq 0.1$. 

\smallskip

\noindent\textbf{Analysis of embedding space:} To evaluate the sensitivity of ISP to the item embedding, we solve $P_{max\_cover@t}$ with several embeddings using \textsc{TextWiser}~\cite{TextWiser}. Besides a baseline Term Frequency Inverse Document Frequency (TFIDF)~\cite{tfidf} embedding, we employ Word2Vec~\cite{word2vec}, GloVe~\cite{pennington2014glove} and Byte-Pair~\cite{sennrich-etal-2016-neural} embeddings to learn word and character level information in the movie and book descriptions. We show that the different embeddings yield similar unit coverage, but that sophisticated NLP methods such as Word2Vec can provide better coverage compared to simpler frequency based embeddings.

\smallskip

\section{Experimental Protocol for ISP + AL}
Given the significant reductions that ISP provides in the time window required for the initial experimentation, our current work is focused on assessing additional benefits of ISP + AL as proposed in Section 5. 

\smallskip 

Ideally, it is desirable to test active exploration in an online recommendation setting. However, the deployment of such a system for testing requires significant effort, and it is also highly undesirable from an end-user perspective. It is therefore mandatory to start with an offline evaluation. Designing an experimental protocol that can closely mimic an online assessment is non-trivial. For that purpose, we propose an offline simulation using the following experimental protocol:

\smallskip 

\noindent{\textbf{Data:}} Given a set of $K$ number of items, we randomly select $k$ items and set those as warm items, referring to the set $M_s$ in Figure~\ref{fig:ISP} and Section~\ref{warm-start}. We set the remaining items as cold items. For all $K$ items, we retrieve their categories and latent embedding as in MovieLens, Goodreads, or other well-known benchmarks.
    
\smallskip 
\noindent{\textbf{Approaches:}} The high-level design given in Figure~\ref{fig:ISP+AL} leaves it open how to combine ISP and AL. The randomized exploration using all $K$ items is a simple baseline method for comparison. Another baseline strategy is to apply the ISP as-is at each round following Day\textendash0. Alternatively, we can also consider the \textit{order} of the selection of cold items. Items selected by ISP earlier in the process, which means they become part of the optimized selection quickly, will have higher weights in the randomized exploration. If we can distinguish between items whose trained models are still unstable, for example, by quantifying the uncertainty in their predictions as shown in~\cite{wang2021modeling}, we can put higher weights to the uncertain models in randomized exploration. 


\smallskip 
\noindent{\textbf{Evaluations:}} 
We repeat the simulation $n$ times with varying sets of $K$ items. For each simulation, we enable exploration for one time period. We then compare how effective the different active exploration strategies are in reducing the number of cold items compared to randomized exploration. This is captured by the total number of models available in the Personalized Recommendations component in Figure~\ref{fig:ISP} and Figure~\ref{fig:ISP+AL}. This is the union of the set of previous warm items, the set of newly added warm items after exploration, and the set of cold items being successfully warm-started by the former two sets. After the exploration, the union of all available models is $M_s \cup M_{s'}$, again shown in both figures. We also calculate the ratio of successful warm-starts by dividing the number of warm-started items by the total number of cold items subject to warm-start, with a higher percentage indicating a more successful strategy.

\bibliographystyle{named}
\bibliography{bibliography.bib}

\begin{thebibliography}{}

\bibitem[\protect\citeauthoryear{Beasley}{1987}]{beasley1987algorithm}
John~E Beasley.
\newblock An algorithm for set covering problem.
\newblock {\em European Journal of Operational Research}, 31(1):85--93, 1987.

\bibitem[\protect\citeauthoryear{Caruana \bgroup \em et al.\egroup
  }{2004}]{DBLP:conf/icml/CaruanaNCK04}
Rich Caruana, Alexandru Niculescu{-}Mizil, Geoff Crew, and Alex Ksikes.
\newblock Ensemble selection from libraries of models.
\newblock In Carla~E. Brodley, editor, {\em Machine Learning, Proceedings of
  the Twenty-first International Conference {(ICML} 2004), Banff, Alberta,
  Canada, July 4-8, 2004}, volume~69 of {\em {ACM} International Conference
  Proceeding Series}. {ACM}, 2004.

\bibitem[\protect\citeauthoryear{Harper and Konstan}{2015}]{movielens2015}
{F. Maxwell} Harper and {Joseph A.} Konstan.
\newblock The movielens datasets: History and context.
\newblock {\em ACM Transactions on Interactive Intelligent Systems}, 5(4),
  December 2015.

\bibitem[\protect\citeauthoryear{Jones}{1972}]{tfidf}
Karen~Sparck Jones.
\newblock A statistical interpretation of term specificity and its application
  in retrieval.
\newblock {\em Journal of documentation}, 1972.

\bibitem[\protect\citeauthoryear{Kadioglu \bgroup \em et al.\egroup
  }{2021}]{kadioglu2021ISP}
Serdar Kadioglu, Bernard Kleynhans, and Xin Wang.
\newblock Optimized item selection to boost exploration for recommender
  systems.
\newblock In {\em Proceedings of the 18th International Conference on
  Integration of Constraint Programming, Artificial Intelligence, and
  Operations Research}, Vienna, Austria, July 2021.

\bibitem[\protect\citeauthoryear{Kilitcioglu and Kadioglu}{2021}]{TextWiser}
Doruk Kilitcioglu and Serdar Kadioglu.
\newblock Representing the unification of text featurization using a
  context-free grammar.
\newblock {\em Proceedings of the AAAI Conference on Artificial Intelligence},
  2021.

\bibitem[\protect\citeauthoryear{Mikolov \bgroup \em et al.\egroup
  }{2013}]{word2vec}
Tomas Mikolov, Ilya Sutskever, Kai Chen, Greg~S Corrado, and Jeff Dean.
\newblock Distributed representations of words and phrases and their
  compositionality.
\newblock In {\em Advances in neural information processing systems}, pages
  3111--3119, 2013.

\bibitem[\protect\citeauthoryear{Pennington \bgroup \em et al.\egroup
  }{2014}]{pennington2014glove}
Jeffrey Pennington, Richard Socher, and Christopher~D. Manning.
\newblock Glove: Global vectors for word representation.
\newblock In {\em Empirical Methods in Natural Language Processing (EMNLP)},
  pages 1532--1543, 2014.

\bibitem[\protect\citeauthoryear{Rubens \bgroup \em et al.\egroup
  }{2015}]{rubens2015active}
Neil Rubens, Mehdi Elahi, Masashi Sugiyama, and Dain Kaplan.
\newblock Active learning in recommender systems.
\newblock In {\em Recommender systems handbook}, pages 809--846. Springer,
  2015.

\bibitem[\protect\citeauthoryear{Sennrich \bgroup \em et al.\egroup
  }{2016}]{sennrich-etal-2016-neural}
Rico Sennrich, Barry Haddow, and Alexandra Birch.
\newblock Neural machine translation of rare words with subword units.
\newblock In {\em Proceedings of the 54th Annual Meeting of the Association for
  Computational Linguistics (Volume 1: Long Papers)}, pages 1715--1725, Berlin,
  Germany, August 2016. Association for Computational Linguistics.

\bibitem[\protect\citeauthoryear{Settles}{2009}]{settles2009active}
Burr Settles.
\newblock Active learning literature survey.
\newblock {\em University of Wisconsin-Madison Department of Computer
  Sciences}, 2009.

\bibitem[\protect\citeauthoryear{Wan and
  McAuley}{2018}]{DBLP:conf/recsys/WanM18}
Mengting Wan and Julian~J. McAuley.
\newblock Item recommendation on monotonic behavior chains.
\newblock In Sole Pera, Michael~D. Ekstrand, Xavier Amatriain, and John
  O'Donovan, editors, {\em Proceedings of the 12th {ACM} Conference on
  Recommender Systems, RecSys 2018, Vancouver, BC, Canada, October 2-7, 2018},
  pages 86--94. {ACM}, 2018.

\bibitem[\protect\citeauthoryear{Wang and Kadioglu}{2021}]{wang2021modeling}
Xin Wang and Serdar Kadioglu.
\newblock Modeling uncertainty to improve personalized recommendations via
  bayesian deep learning.
\newblock {\em International Journal of Data Science and Analytics}, pages
  2364--4168, 2021.

\end{thebibliography}

\end{document}